\documentclass[vecphys]{svmult}

\usepackage{graphicx}        
\usepackage{multicol}        
\usepackage[bottom]{footmisc}

\begin{document}

\title*{Laboratory Studies of Astrophysical Jets}
\author{Andrea Ciardi}
\institute{Observatoire de Paris, LERMA, 5 Place Jules Janssen, 92195 Meudon, France
\texttt{andrea.ciardi@obspm.fr}}
%
%
\maketitle

\section{Introduction}
\label{sec:introduction}
The connection between astrophysics and laboratory experiments is often made in the context of atomic physics, and in particular spectroscopic studies. These generally concentrate on the micro-physics of astrophysical plasmas, providing important data on cross-sections, opacities, grain chemistry, and other processes that are vital to the modelling and interpretation of astronomical observations. Indeed, detailed laboratory opacity measurements of iron absorption lines have greatly improved models of pulsation periods in Cepheid variables \cite{daSilva92}.

Recreating large-scale astrophysical phenomena in the laboratory has also inspired scientists. A pioneer in the use of laboratory experiments to study space physics was the Norwegian scientist Kristian Birkeland (1867-1917), who used a terrella\footnote{A terrella is a small (tens of cm) magnetized model Earth.} and gas discharges to investigate auroral phenomena \cite{Egeland05}. Indeed, as early as the 17th century a terrella was used by William Gilbert to study the Earth's magnetism \cite{Mettelay1893}. More recently, the idea that physical problems in the cosmos may be elucidated by more mundane earth-based events, such as fluid flows and nowadays laboratory plasmas, was taken up in a series of meetings held just after World War II, with the aim: ``\textit{To bring together workers from astrophysics and from aerodynamics; ...to consider which developments in fluid mechanics may be applicable to astrophysical problems, and to arrive at a formulation of these problems in such a way that mathematicians and fluid mechanics people may find a way of attack}'' \cite{van-de-Hulst55}.
Topics of discussion included accretion, gas and dust in the interstellar medium, shocks, and turbulence; the latter being very popular with those working in fluid mechanics. An indication that this was taken rather seriously can be gleaned from the list of scientist who took part in the meetings. These included the likes of Heisenberg, Sciama, Bondi, Hoyle, Bok, Burgers, Seaton and Cowling. At the meeting held in Cambridge in 1954, Chandrasekar and Fermi contributed a preliminary paper but did not actually attend; while Hubble, although listed as participant, had died in fact nine months earlier\footnote{As a final piece of trivia, Hubble's funeral was never held and the location of his resting body never disclosed.}. 

In the last ten years or so, experiments on pulsed-power facilities (z-pinches) and high-power lasers are leading the way in studying astrophysical phenomena in the laboratory. Emerging areas of research have been aimed at producing complex dynamical phenomena, such as compressible hydrodynamic mixing, hypersonic jets, shock physics, radiation hydrodynamics and photoionised plasmas, to name a few. These can help to understand the physics of phenomena associated with wide range of astrophysical objects, including protostellar and AGNs jets, supernovae explosions and the subsequent generation of remnants, and photoevaporated molecular clouds. Here we will restrict the discussion to the jets produced on z-pinch facilities and in particular to the work performed on the MAGPIE generator at Imperial College. For a broader and more detailed discussion of laboratory astrophysics on lasers and z-pinches we refer the reader to the review by Remington et al. 2006\cite{Remington06} and the book by Drake 2006 \cite{Drake06}.

\section{Plasma conditions in z-pinch and laser experiments}
\label{sec:2}
%

\begin{figure}
\centering
\includegraphics[width=11.5cm]{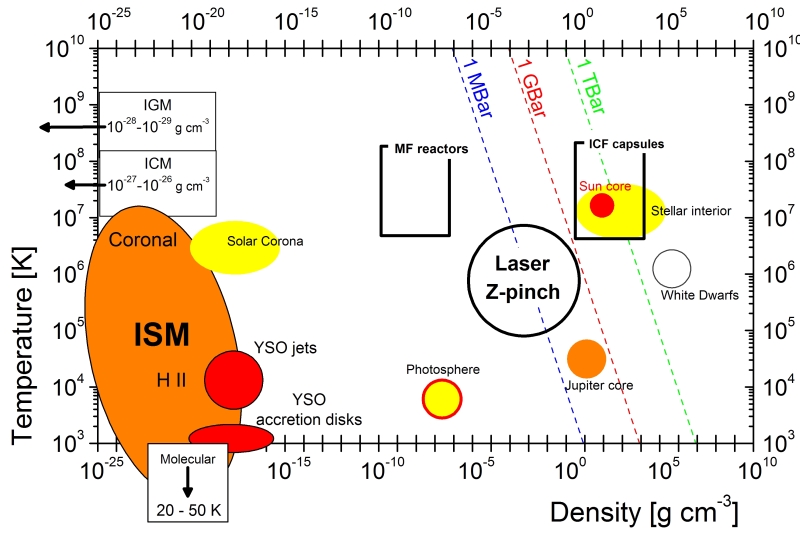}
%
%
\caption{Plot of the temperature versus density for a variety of laboratory and astrophysical plasmas. Lines of constant pressure are shown for fully ionized hydrogen. High-energy density laboratory astrophysics is in the regime of pressures ${\geq}$ 1 Mbar and the typical conditions currently obtained in laser and z-pinch experiments are easily in this range. The conditions that will be accessible in future Magnetic Fusion (MF) reactors and Inertial Confinement Fusion (ICF) laser experiments are also indicated. The main phases of the Interstellar Medium (ISM) are shown. The Intergalactic Medium (IGM) and the Intracluster Medium (ICM) lie outside the plot at lower densities.}
\label{fig:1}
\end{figure}

Typical plasmas produced on z-pinch and laser facilities have pressures of ${\sim}$ Mbar, corresponding to energy densities ${\sim 10^{12}}$ erg cm${^{-3}}$, at a fraction of solid density. An overview of the plasma conditions attainable on experimental installations, together with some of those found in space is given in Fig. \ref{fig:1}. Z-pinch facilities rely on stored electrical energy (hundreds of kJ) to deliver large currents (${\sim}$ of a few Mega Amperes) over a short time (${\sim 100 - 1000}$ ns) to a ``load'' usually consisting of a gas or thin metallic wires. These facilities typically produce volumes of plasma ${\sim 1}$ cm${^{-3}}$ (for a review see \cite{Ryutov00a}). Laser facilities instead rely on focusing onto a solid or gaseous target, single or multiple high-power laser beams. These concentrate several kJ of energy, over timescales $\sim$ pico- to nano-second, into plasma volumes of ${\sim 1}$ mm${^{-3}}$. The future arrival of the Laser Megajoule (LMJ) facility in France and the National Ignition Facility (NIF) in the USA, will produce fusion plasmas under conditions similar to stellar interiors.
 
For the present discussion (i.e. jets) it is more interesting to look at the dynamical conditions that can be obtained in the laboratory. When the energy available on z-pinches and lasers is partially converted into kinetic energy, it can generate hypersonic (Mach numbers  ${M>5}$), radiatively cooled flows with characteristic velocities of the order of 100 - 1000 km s${^{-1}}$. These flows can include dynamically important magnetic fields, $\sim$ several 10${^6}$ Gauss, and have a large range of plasma-${\beta}$ (the ratio of thermal to magnetic pressure), ${1 >> \beta >> 1}$. In such cases, our inability to obtain the adequate astrophysical plasma conditions (see Fig. \ref{fig:1}) may be overcome by producing scaled ``conditions'' of the phenomena of interest. These are discussed in the next section.

\section{Relating laboratory and astrophysical phenomena}
\label{sec:3}
The framework to relate experiments to astrophysical phenomena, within magnetohydrodynamics, was developed in a series of papers by Ryutov et al. \cite{Ryutov99,Ryutov00b,Ryutov02,Ryutov03}. A general discussion of hydrodynamic scaling can also be found in \cite{Zel'Dovich67}, and in the present volume it is reviewed in detail by Cocker. An interesting duality between imploding and exploding systems is given in \cite{Drury00}. For collisionless plasmas the scaling conditions are descibed in \cite{Drake00}.
 
\begin{figure}
\centering
\includegraphics[width=11.5cm]{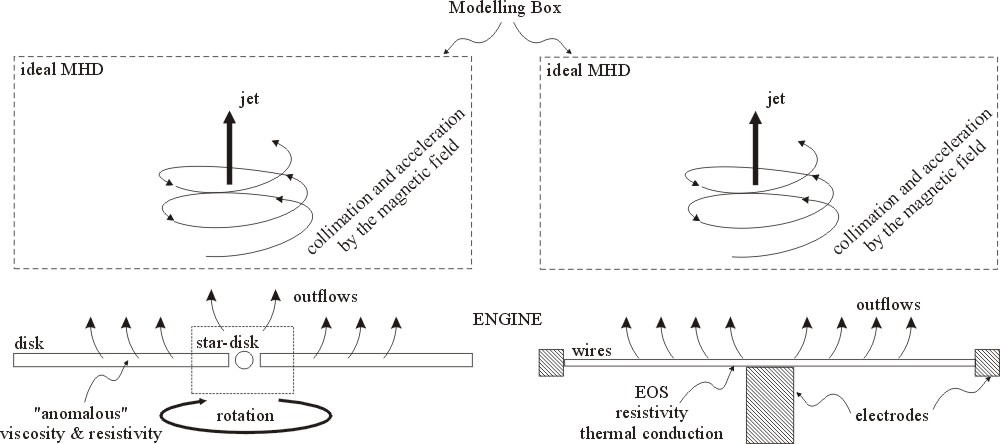}
%
%
\caption{The lower part of each plot shows schematically the ``engine'' that produces astrophysical (left) and laboratory jets (right). The physical mechanisms are clearly very different, however we are interested in the scaling of the flows inside the Modelling Boxes.}
\label{fig:2}
\end{figure}

	Here we will qualitatively introduce some of the ideas behind scaling and shall be concerned with systems which are to a good degree magnetofluids. As a specific example we take the MHD jet launching, which will be later discussed in more details in the context of laboratory experiments. Fig. \ref{fig:2} shows schematically the comparison of the ``environment'' producing astrophysical and laboratory jets. The details of the experimental set-up depicted are not important at this stage, it is sufficient to say that the physical processes leading to the distributions of magnetic fields, plasma density, pressure and velocity inside the ``modelling box'' (MB) are completely different for the two systems. However because we are interested in studying experimentally the collimation and launching of the jet in the MB, we are concerned only with the differences that may be present there. The physical model to be employed in the MB is that of ideal MHD (for an introduction to MHD see \cite{Pelletier07}). While this is a good approximation for the astrophysical case, it is not usually the case for the experiments, which have to deal with non-ideal effects arising for example from thermal conduction, viscosity and resistivity. Nevertheless the experiments can be \textit{designed} so that the ideal MHD approximation is valid at least inside the MB, thus making the set of equations describing the plasma evolution in the two MB the same. For scaling purposes this is however not sufficient, it is also fundamental for the initial (and boundary) conditions in the two MB to be geometrically similar. Meaning that at some arbitrary time, the initial spatial distributions of all physical quantities (density ${\rho}$, velocity ${\textbf{V}}$, pressure ${P}$ and magnetic field ${\textbf{B}}$) are the same. Finally, the actual scaling factors relating the physical variable in the two systems and their initial distributions should obey the following constraints:
\[
\tilde{V_L}\sqrt{\frac{\tilde{\rho_L}}{\tilde{P_L}}}=\tilde{V_A}\sqrt{\frac{\tilde{\rho_A}}{\tilde{P_A}}}, ~~	
\frac{\tilde{B_L}}{\sqrt{\tilde{P_L}}}=\frac{\tilde{B_A}}{\sqrt{\tilde{P_A}}}
\]
where the tilde denotes characteristic (dimensional) values for the laboratory (L) and astrophysical (A) plasmas. If the aforementioned conditions are met, namely the ideal MHD equations hold, the initial conditions are geometrically similar and the constraints on the scaling factors hold, then the evolution in the modelling boxes, from some initial time and over the scaled time span, will be \textit{indistinguishable}. The addition of extra physical processes, for example radiation, increases the constraints to be satisfied and hence the difficulty in obtaining an exact scaling (see for example \cite{Castor07}).

Two points need to be clarified. The first regards obtaining geometrically similar initial conditions in the laboratory. This is clearly very difficult and it has been done exactly in only a few cases\cite{Remington97}. The question also concerns which initial condition should be used. Obviously observations cannot be expected to provide three-dimensional distributions of all the quantities of interest, if they did the problem would be almost solved! Then, with observational constraints in mind, we resort to models, which are in general as difficult to reproduce. The usefulness of laboratory astrophysics however goes beyond strict scaling. By using ``reasonable'' initial conditions, as it is usually done in numerical simulations, we can study in detail complex astrophysical phenomena in a repeatable and accessible manner. In addition the possibility of validating astrophysical codes using laboratory data is also very important. The second point is on the accuracy of the approximation of ideal MHD for laboratory plasmas, and more specifically the ones produced on z-pinch installations. Neglecting the effects associated with viscosity, thermal conduction and resistivity (the ideal MHD approximation) means that the Reynolds (Re), Peclet (Pe) and magnetic Reynolds (Re${_m}$) numbers are Re, Pe, Re${_m} >> 1$. While for YSO jets these dimensionless numbers are very large (see Table \ref{tab:1}), for laboratory jets they are generally many orders of magnitude smaller, thus calling into question the validity of the ideal MHD approximation. However the values obtained in the experiments should be compared to astrophysical numerical simulations, where because of the finite accuracy of the numerical schemes employed, the ideal MHD approximation also breaks down. Unless included explicitly in the simulations, viscous and resistive dissipation, and thermal conduction occur at the grid level through unphysical numerical truncation errors. Existing ideal MHD simulations of jets have dimensionless numbers in the range ${50-1000}$, well within the reach of laboratory experiments. In some case, as for the Reynolds number, the laboratory values can be orders of magnitude larger. Thus laboratory astrophysics experiments can be very useful in studying complex and intrinsically non-linear three-dimensional phenomena, in regimes where accurate numerical simulations are particularly difficult to perform, and thus it represents a powerful complementary tool to astrophysical modelling. As a final justification: ``Real experiments are also irreplaceable in providing new insights into subtle physics issues and in stirring the creative imagination of scientists''\cite{Ryutov02}.

\section{Young stellar jets from Z-pinch machines}
\label{sec:4}
In this section we shall discuss the z-pinch studies of astrophysical jets. It is worth mentioning that the time for an experiment to reach ``maturity'': from the conception, design and first experiments to the modelling, data analysis and then the application to astrophysical models is of the order of 3 to 5 years, depending on its complexity. Considering that most of the experiments, both on laser and on z-pinches, are less than 10 years old, it should not come as a surprise that in many instances the applications to astrophysics are not yet fully developed or very clear. Although we shall restrict our discussion to jets, there are other areas of laboratory astrophysics research on z-pinches that are well developed, such as equations of state studies for planetary interiors and photoionised plasmas relevant to accretion disks around compact objects. These are thoroughly reviewed in \cite{Remington06} and we refer the interested reader to that paper.

\subsection{Hydrodynamic jets from conical wire arrays}
\label{sec:4.1}
\begin{figure}
\centering
\includegraphics[width=11.5cm]{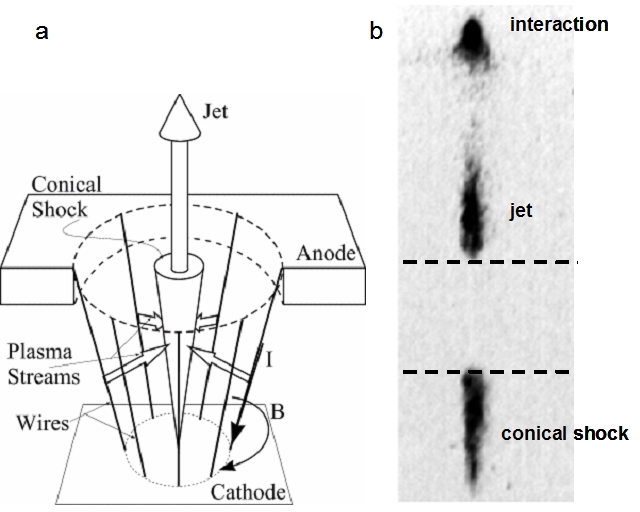}
%
%
\caption{(a) Typical arrays are made with 16 tungsten wires, each with a diameter of 18 $\mu$m. The smaller radius of the array is 8 mm and the wires are inclined at an angle of 30$^\circ$ with respect to the axis. The axial length of the array is 12 mm. Continuous plasma streams converge on axis producing a conical shock which redirects the flow axially. (b) Time-resolved, filtered XUV emission from a laboratory jet. Emission from the region in between the dotted lines is screened by the anode. As the jet propagates it cools down and its emission decays, however it is again visible where the jet interacts with a background plasma.}
\label{fig:3}
\end{figure}

Conically converging flows were investigated in astrophysics as a possible mechanism for converting wide angle winds into collimated jets. Such models do not require magnetic fields, at least to collimate and launch the jet, and rely instead on purely hydrodynamic means\cite{Tenorio-Tagle88,Canto88}. Within this framework, and with the aim of producing hydrodynamic jets to be used for interaction studies, a series of experiments\cite{Lebedev02,Ciardi02} were developed on the z-pinch generator MAGPIE. The schematic of the experimental configuration is shown in Fig. \ref{fig:3}. It consists of a conical array of micron-sized metallic wires driven by a current of 1 MA rising to its peak value in 240 ns. The basic mechanism of plasma formation in wire arrays is the following: resistive heating rapidly converts the wires into a heterogeneous structure consisting of a cold ($<1$ eV) and dense liquid-vapour core, surrounded by a relatively hot ($10-20$ eV) and low density (${\sim}$10${^{17}}$ cm${^{-3}}$) plasma. Most of the current flows in the latter, where the resistivity is lower, which undergoes acceleration by the $\textbf{J}\times\textbf{B}$ force toward the array axis. These streams of plasma have characteristic velocities of ${\sim100-150}$ km s${^{-1}}$ and corresponding Mach numbers ${M\sim5}$. The wire cores act as a reservoir of plasma, replenishing the streams during the entire duration of the experiment (several hundred ns). The converging plasma is virtually magnetically field-free and the interaction on axis is hydrodynamic in character. The collision produces a standing conical shock where part of the kinetic energy of the streams is thermalized. However it is important to note that the plasma streams are not perpendicular to the surface of the conical shock. Thus the component of the velocity parallel to the shock is continuous across the shock and the flow is redirected upwards into a jet. Typical jet velocities are ${\sim100-200}$ km s${^{-1}}$ and hypersonic jets with ${M>10}$ can be produced by this mechanism. The jet collimation and Mach numbers depend predominantly on the level of radiation cooling in the plasma, which can be changed experimentally by varying the wire material (Al, Fe, W and so on). Increasing the atomic number of the wires increases the rate of cooling from the plasma, lowers its temperature and leads to the formation of more collimated jets (with higher Mach numbers) \cite{Ciardi02,Lebedev02,Lebedev05a}. These jets are used to study the propagation and interaction with an ambient medium, which are described in sec. \ref{sec:4.3}. The characteristic conditions and dimensionless parameters obtained are shown in Table \ref{tab:1}.

\begin{figure}
\centering
\includegraphics[width=11.5cm]{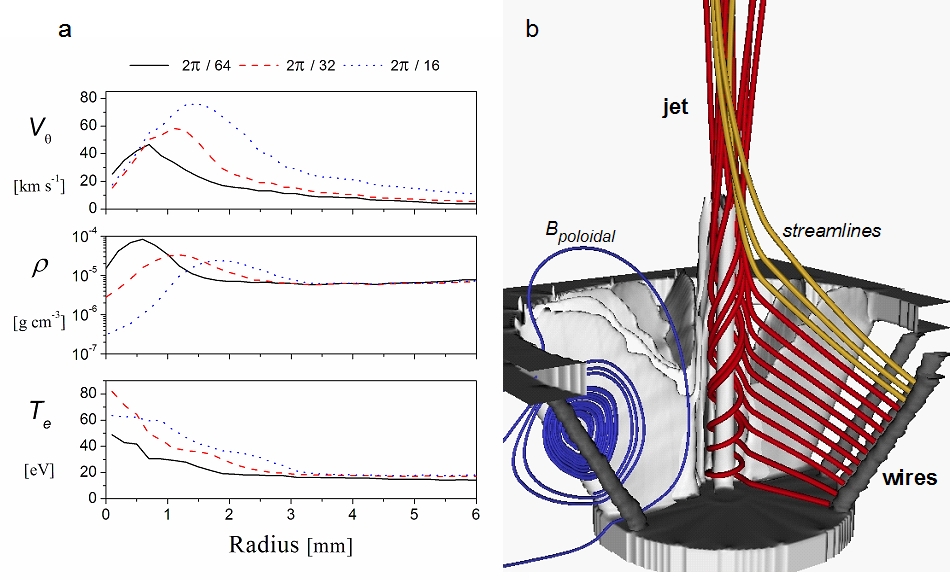}
%
%
\caption{The experimental set-up is similar to that shown in Fig. \ref{fig:3}, in this case however the wires are twisted in the azimuthal direction (see main text) (a) Azimuthally averaged profiles of azimuthal velocity, density and electron temperature as a function of radial position, taken 6 mm above the cathode 200 ns after the start of the current pulse. The profiles are for different twist angles of the array, which can be used to modify the angular rotation in the jet. (b) Isodensity surfaces at 270 ns show the dense plasma around the wires (10$^{-3}$ g cm$^{^3}$, dark gray) and plasma streams ($2.5\times10^{-5}$ g cm$^{^3}$, light gray); velocity streamlines from one of the wires are shown with the oppositely rotating flows separated visually by red and orange streamlines. The oppositely rotating flow is due to electrode's effects and it is dynamically unimportant. The azimuthally averaged poloidal magnetic field lines are shown in blue.\cite{Ample08}}
\label{fig:4}
\end{figure}

It is possible to design and modify the experiments to include additional physics, such as dynamically dominant magnetic fields and rotation. Indeed, for accretion onto the forming star to occur, angular momentum needs to be removed from the in-falling material. In combination with the processes present in the accretion disk, such as MHD instabilities and turbulent transport, jets and winds can also remove a considerable fraction of the excess angular momentum from the accreting flow \cite{Combet08}. One of the obvious implications is that jet will be rotating and some confirmation has arrived with recent observations of rotation in a number of YSO jets \cite{Coffey04}. Supersonically rotating laboratory jets and flows of astrophysical relevance were recently produced for the first time \cite{Ample08} using a variant of the conical wire array. Rotation in the flow is accomplished by slightly twisting the wires in the azimuthal direction. This results in a poloidal magnetic field and an azimuthal component of the Lorentz force, giving a non-zero torque on the plasma streams (Fig. \ref{fig:4}). The level of angular momentum introduced in the system can be controlled by changing the twist angle and in general the jets ejected have rotation velocities ${\sim100-200}$ km s${^{-1}}$, corresponding to ${\sim0.1-0.2}$ of the jet propagation velocity. One of the applications of these proof-of-principle experiments will be to study the effects of rotation on the propagation of jets and on the growth of the Rayleigh-Taylor instability in curved jets (see section \ref{sec:4.3})

\subsection{Magnetohydrodynamic jets} 
Protostellar (and galactic jets) are thought to be powered by the combination of rotation and magnetic fields, which extract the rotational energy from an accreting system and create magnetic stresses which accelerate and collimate the flow (see the lectures notes \cite{Ferreira07,Tsinganos07}. Depending on the details of the models, the winding of an initially poloidal magnetic field results in a flow pattern dominated by a toroidal field. A similar situation is also attained when the foot-points of a field line, connecting the disc to a central compact object or connecting different parts of a disc, rotate with different angular velocities. In such cases, the relative angular displacement of the foot-points causes one of them to move ahead of the other, and the field loop to twist. The induced toroidal component results in an increase of the magnetic pressure which drives the expansion of the loop itself \cite{Lovelace95}. In the magnetic tower scenario \cite{Lynden-Bell06,Lynden-Bell03}, the outcome is a magnetic cavity consisting of a highly wound up toroidal field which accelerates the flow. In this case, the presence of an external plasma medium was shown to be necessary to confine the magnetic cavity, which would otherwise splay out to infinity within a few rotations \cite{Lynden-Bell96}. The basic picture of magnetic tower evolution has also been confirmed numerically by several authors \cite{Kato04a,Kato04b,Nakamura06,Nakamura07,Matt06}.

\begin{figure}
\centering
\includegraphics[width=11.5cm]{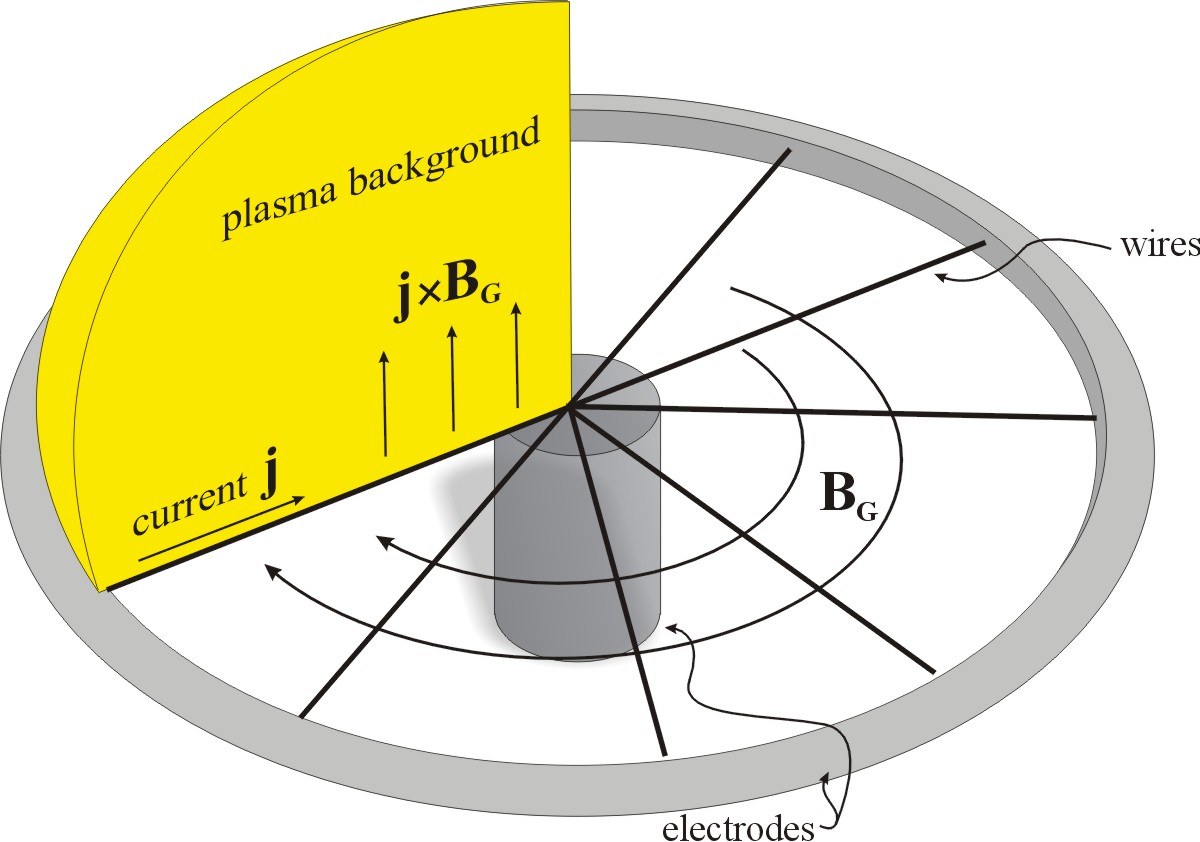}
%
%
\caption{Schematic of a radial wire array.}
\label{fig:5}
\end{figure}

\begin{figure}
\centering
\includegraphics[width=11cm]{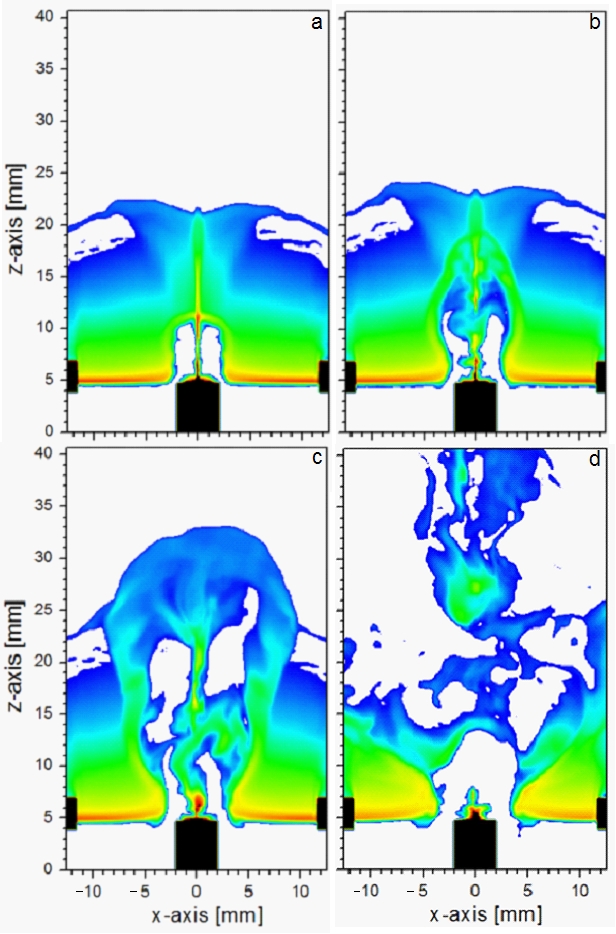}
%
%
\caption{Slices of mass density from a 3D MHD simulation of radial wire arrays. The evolution is shown at times (a) 165 ns, (b) 175 ns, (c) 185 ns and (d) 205 ns. The logarithmic density scale is from 10$^{-7}$ g cm$^{^3}$ (blue) to 10$^{-1}$ g cm$^{^3}$ (red). Regions in white are void of plasma, but not electromagnetic fields, and are essentially a computational ``vacuum``. The square, black regions are the electrodes.\cite{Ciardi07a}}
\label{fig:6}
\end{figure}

\begin{figure}
\centering
\includegraphics[width=11.5cm]{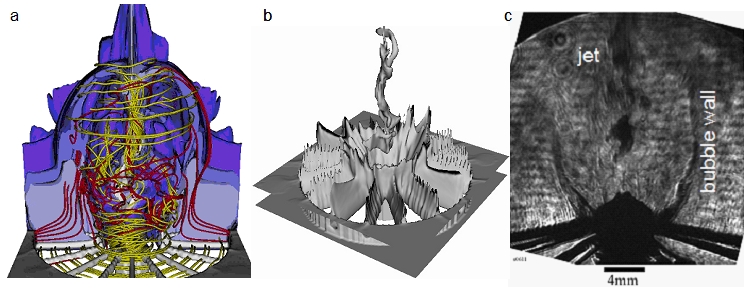}
%
%
\caption{(a) Magnetic field lines (yellow) and current density (red) distribution inside the magnetic cavity at 245 ns \cite{Ciardi07a}. To show the inside of the magnetic cavity the isodensity contours (shades of blue) are sliced vertically. With the onset of the kink instability the magnetic field wraps tightly around the jet, which is seen more clearly in (b) without the cavity walls\cite{Ciardi07b}. (c) Experimental shadowgraph showing the bubble wall and the clumpy jet launched along the axis\cite{Lebedev05b}.}
\label{fig:7}
\end{figure}

\paragraph{Magnetically driven jets from radial wire arrays}
The study of magnetically collimated and accelerated jets on z-pinches was developed in the last few years \cite{Ciardi05,Lebedev05b,Ciardi07a,Ciardi07b} using a modified wire array configuration. The basic astrophysical mechanism studied in the experiments is the interaction of a toroidal magnetic field with a plasma ambient medium, leading to the formation of jets and magnetic ``bubbles''. The schematic of the experimental set-up, a radial wire array, is shown in Fig. \ref{fig:5}. The formation of  plasma is similar to that discussed in conical wire arrays, however the plasma is now accelerated vertically filling the space (few cm) above the array. Below the wires there is only a toroidal magnetic field. The formation of the jet and its time evolution is shown in Fig. \ref{fig:6}.  The initial formation of the magnetic cavity and jet occurs at the time when the magnetic pressure is large enough to break through the wires. This occurs only over a small region close to the central electrode, where the toroidal magnetic field ${B_G}$ is strongest. The results show the system evolving into a structure consisting of an approximately cylindrical magnetic cavity with an embedded jet on its axis confined by the magnetic ``pinching'' force. A shell of swept-up plasma surrounds and partially confines the magnetic bubble. The subsequent evolution is dominated by current-driven instabilities and the development of the asymmetric ``kink'' mode (${m = 1}$) which leads to a distortion of the jet and a re-arrangement of the magnetic field. In Fig. \ref{fig:7}a, the magnetic field lines can be seen to twist inside the jet, an effect caused by the instability which turns toroidal into poloidal magnetic flux. The end result of the instabilities however is not to destroy the jet, but to produce an inhomogeneous or ``knotty'' jet, shown in Fig. \ref{fig:7}b-c. The resulting jet has typical super-fast-magnetosonic Mach numbers in excess of 5, it is kinetically dominated and its opening angle $< 20{^{\circ}}$.

The relatively simple initial conditions implemented experimentally produce a very complex and rich dynamics which share many important features with astrophysical models. One important example is the presence of a envelope surrounding the magnetic cavity and confining it. Although this is discussed in the astrophysical literature \cite{Lynden-Bell06,Uzdensky06}, it has so far only been observed in a laboratory experiment. The stability and dynamics of the envelope, which determine the collimation of the cavity itself, can thus be directly studied in the laboratory before astronomical observations may become available. Finally, it is worth pointing out that while two-dimensional, axisymmetric MHD simulations reproduce very well the experimental results, up to the development of the non-asymmetric current-driven instabilities. There are fundamental differences in the long-term evolution of the system, which can only be reproduced by fully three-dimensional simulations. 

\paragraph{Episodic ejection of magnetic bubbles and jets}
Protostellar jets are characterized by the presence of knots and multiple bow-shocks \cite{Hartigan05}, tracing their propagation. These are often interpreted as internal shocks driven by relatively small perturbations in a steady ejection process, and which occur on typical time-scale between $\sim5-20$ years. For example in \cite{deColle08} it was shown that the temporal variability of the jet velocity may be associated with a time-varying stellar magnetic field. Episodic jet ejection behaviour may also be associated with variation in the accretion rates or an inflating stellar magnetosphere.

\begin{figure}
\centering
\includegraphics[width=11.5cm]{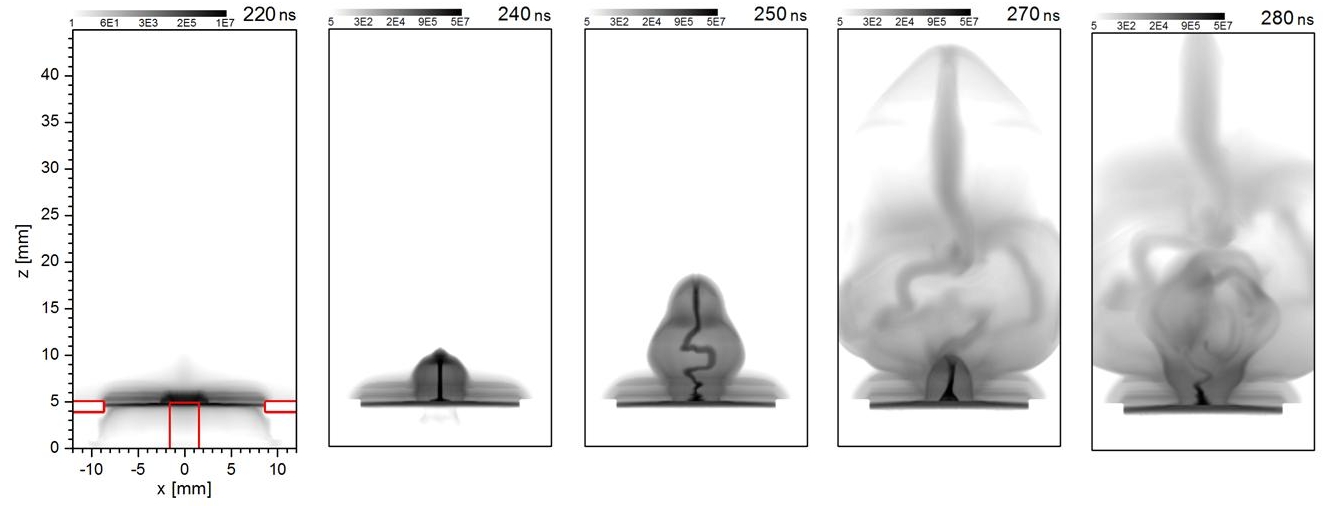}
%
%
\caption{3D MHD simulation of a radial foil experiment. The images show the line-of-sight integrated emission (in arbitrary units) at different times. The red line in the first panel shows the position of the electrodes.}
\label{fig:7b}
\end{figure}

\begin{figure}
\centering
\includegraphics[width=11.5cm]{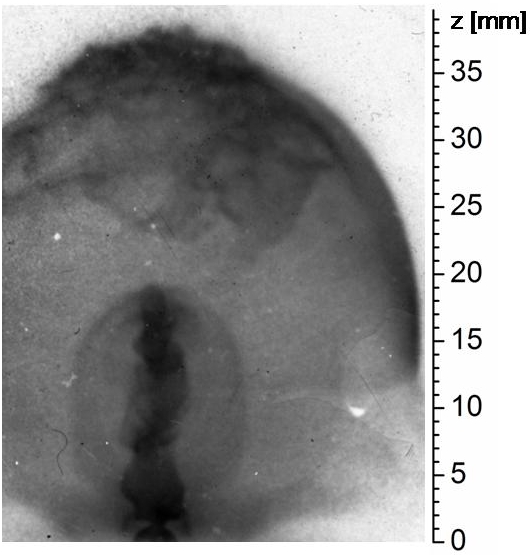}
%
%
\caption{Experimental image showing the presence of nested magnetic cavities with an embedded jet. The image shows the self-emission in the XUV range \cite{Ciardi08b}}
\label{fig:7c}
\end{figure}

Recent experiments have studied for the first time the episodic ejection of magnetic bubbles and jets, and its effects on the overall propagation of the outflow\cite{Ciardi08b}. The experimental set-up is similar to that shown in Fig. \ref{fig:4}, however the wires are replaced by a 6 $\mu$m thick metallic foil (usually aluminium). A 3D MHD simulation of the experiments is shown in Fig. \ref{fig:7b}. The evolution of the first bubble is similar to that of radial wire arrays. However the total mass in the plasma source, as a function of radius, is larger for a foil than for a radial wire array. Thus after the first magnetic cavity and jet are formed, there is a larger quantity of plasma available to refill the ``gap''  between the central electrode and the left over foil; for example the presence of this gap is visible in \ref{fig:5} for radial arrays, and it is produced by the magnetic field pressure breaking through the wires or foil. Once the gap is refilled with plasma, the currents can flow once again across the base of the magnetic cavity, thus re-establishing the initial configuration. When the magnetic pressure is large enough to break through this newly deposited mass, a new jet/bubble ejection cycle can begin. Typical flow velocities observed are $\sim100-400$ km s${^{-1}}$, the simulated sonic and the alfv\'{e}nic Mach numbers in the jet, defined as the ratios of the flow speed to the sound and Alfv\'{e}n speed respectively, are $M_s \sim M_A \sim 3-10$. The resulting flow is heterogeneous and clumpy, and it is injected into a long lasting and well collimated channel made of nested cavities. Each jet/outflow episode propagates, interacts and substantially alters the surrounding environment by injecting mass, momentum, energy and magnetic flux into it. An important aspect of the episodic ejection process is, broadly speaking, its self-collimation. Since the initial ambient medium is swept away after a few ejections, newly formed magnetic cavities are confined solely by the environment left by earlier episodes, thus making the collimation process insensitive to the initial ambient conditions. An experimental image of the evolution of the system is shown in Fig.\ref{fig:7c} In the magnetic cavities $Re_M > 100$, and each bubble expands with its own ``frozen-in'' magnetic flux; in the experiments this is confirmed by the magnetic probe measurements of the trapped magnetic field at the outer edge of the bubbles, $B\sim1-5$ kG. The collimation is then determined not only by the pressure of the left-over plasma but also by the pressure of the tangled magnetic field trapped in the bubbles, where the plasma-$\beta$ is in the range $0.1<\beta<1$. A high level of symmetry is maintained after many ejections (5 in the current experiments), the number being limited only by the duration of the current pulse delivered by the generator. Overall, the experiments demonstrate that magnetic acceleration and collimation, occurring within a framework of strongly episodic outflow activity, can be effective in producing well collimated and heterogeneous jets.

By drawing a parallel with the dynamics observed in the experiments, one can gain useful insights and a qualitative view of the possible evolution of astrophysical jets. In the experiments there are two time-scales which determine the magnetic bubbles/jets development:  the current-driven (CD) instability time-scale $\tau_I$ and the episodic bubble ejection time-scale $\tau_B$. For conditions applicable to the formation region of protostellar jets \cite{Hartigan07} we can estimate the growth time of the CD kink mode as the Alfv\'{e}n crossing time $\tau_I\sim1$ year; corresponding to a few nanoseconds in the experiments. The second time-scale is linked to the temporal variability of the Poynting flux feeding the bubbles, and for astrophysical sources $\tau_B$ should be associated with a substantial variation in the outflow launching activity; observations of knots kinematics suggest characteristic times $\tau_B\sim5-20$ years; the experiments are in a similar regime  . Because both time-scales and are relatively longer than the characteristic Keplerian period of rotation at the inner disk radius, jet launching should have ample time to reach steady-state. The characteristic astrophysical flow velocities can be taken to be $v\sim200$ km s$^{-1}$. With these conditions, the presence of multiple bubble-like features should be observed on scales   ranging from a few tens to a few hundred AU from the source. Indeed ejection variability, limb-brightened bubble-like structures and the presence of wiggles in the optical DG Tau jet are evident on scales ranging from of a few tens to a few hundred AU the source \cite{Bacciotti00,Dugados00}. The experiments also indicate that asymmetries in the flow can be produced by instabilities that do not destroy the collimation, and because of their relatively short growth time  , jets should develop non-axisymmetric features already within a few tens AU from the source, and become more heterogeneous and clumpy as they move further away to hundreds of AU. It was recently reported for a number of TTauri jets, including DG Tau, that already within 100 AU from the source the jet physical conditions show considerable asymmetries with respect to the axis \cite{Coffey08}. Finally over the same length scales the experiments suggest magnetic energy dissipation, heating of the plasma and a transition to a kinetically dominated jet which propagates ballistically. X-ray emission from the DG Tau jet was recently detected on the same length scales and it was proposed that magnetic energy dissipation may be behind the heating mechanism \cite{Gudel08}. As in the experiments, instabilities and the tangling of the magnetic field may provide a compelling route to efficient heating of such plasmas.
	
\subsection{Interaction with the interstellar medium}
\label{sec:4.3}
\paragraph{Curved jets}
A number of bipolar Herbig-Haro (HH) jets exhibit a distinguishing C-shape morphology indicative of a steady bending \cite{Bally01}. Less regular curvature is also observed in a number of other HH jets; for example in HH 30 a small side drift close to the jet source is followed further away by a sudden bending \cite{Anglada07}. In general, the curvature in jets has been linked either with the motion of the jet sources relative to the ambient medium or with the presence of a widespread outflow; both cases giving rise to an effective transverse ram pressure (cross-wind) which curves the jet. Expected wind velocities vary from a few km s$^{-1}$ for the jet-wind interaction associated with relative motions of TTauri stars with respect to the surrounding environment, (see for example \cite{Jones79}) to typically higher velocities for irradiated jets, where best fits to HH505 H$\alpha$ emission maps were obtained for a wind velocity of 15 km s${^{-1}}$\cite{Masciadri01} and estimates in \cite{Bally01} give wind velocities in the Orion nebula and in NGC1333 of ${\sim10-20}$ km s${^{-1}}$.

\begin{figure}
\centering
\includegraphics[width=11.5cm]{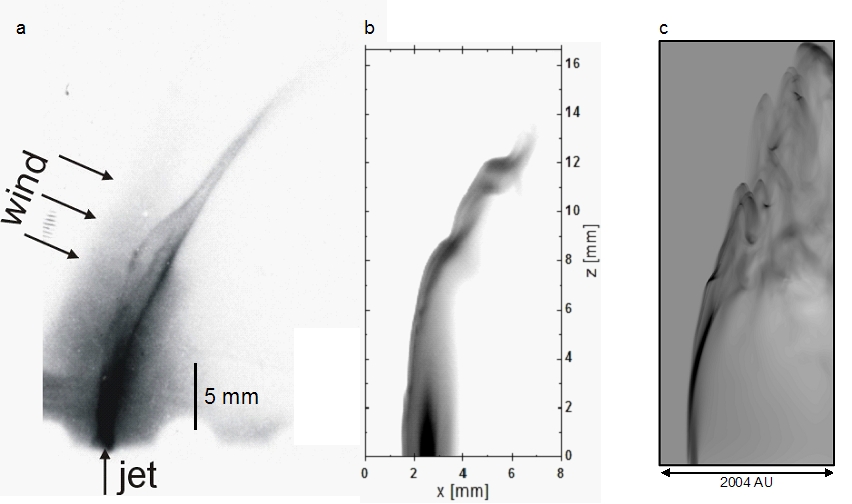}
%
%
\caption{(a) Experimental time-resolved XUV image of a curved jet. (b) Synthetic XUV image of a curved laboratory jet from a 3D simulation. (c) Column density from a 3D simulation of an astrophysical jet\cite{Ciardi08}}
\label{fig:8}
\end{figure}

The hydrodynamic laboratory jets described in section \ref{sec:4.1} are ideally suited to study the interaction with an ambient medium. The region into which the jet is launched (cf. Fig. \ref{fig:3}) is a large vacuum that can be easily filled with different types of background gases. To investigate the dynamics of curved HH jets, a cross-wind was produced by a radiatively ablated foil appropriately placed in the jet propagation region \cite{Ample07,Lebedev05a,Lebedev04}. Typical wind velocities ${\sim30-50}$ km s${^{-1}}$ can be produced in the laboratory, with the important parameters characterising the interaction in the range $V_{jet}/V_{wind}\sim2-4$  and $n_{jet}/n_{wind}\sim0.1-10$. Fig. \ref{fig:8} shows an example of the experiments and simulations of curved jets\cite{Ciardi08}. The characteristic dynamics of the interaction is similar for the laboratory and astrophysical systems, showing notably the formation of new working surface in the jet and ``knotty'' structure in the flow. Curved jets are also Rayleigh-Taylor (RT) unstable, with the growth of such mode disrupting their propagation. However it was shown that jet rotation may partially suppress the instability by shearing the RT modes and confining the perturbations to a narrower layer of the jet body. Nevertheless this promotes the development of the Kelvin-Helmholtz instability (at least for the subsonically rotating jets) which is later responsible for disrupting the jet. Experimentally the RT growth time is of the order of the dynamical time over which the interaction can be produced and new, longer time-scale experiments will be needed to observe its full development.

\begin{figure}
\centering
\includegraphics[width=11.5cm]{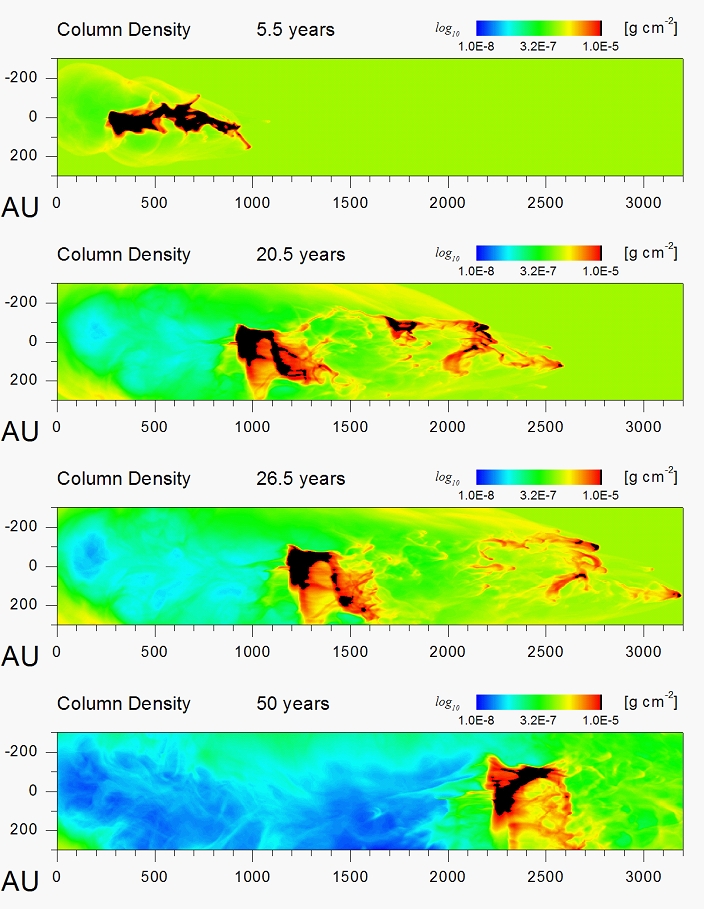}
%
%
\caption{Propagation of an episodic protostellar jet showing the break up into small and large clumps.}
\label{fig:9}
\end{figure}

\paragraph{Clump propagation}
We now return to the scaling issue of some of the laboratory flows and in particular the MHD jets. As we have seen their  evolution is dominated by current-driven instabilities, and the resulting flow is inherently time-dependent and inhomogeneous. To study the propagation of such flows in an astrophysical setting, the data obtained from laboratory MHD jet simulations can be scaled-up and used as initial conditions to model astrophysical clumpy jets. There is clearly some arbitrariness on the choice of some of the scaling parameters and for the case presented here we assume the flow to be close to the YSO source. Noting that the laboratory and astrophysical jet velocities are of the same order, we choose the following three scaling: $V_{lab}=V_{astro}$, $L_{lab}=1$ mm ${\rightarrow}$ $L_{astro}=10$ AU and $\rho_{lab}=10{^{-3}}$ g cm${^{-3}}$ ${\rightarrow}$ $\rho_{astro}=10{^{-18}}$ g cm${^{-3}}$. The choice of scale-length gives an initial jet radius ${\sim}$20 - 30 AU and we take the maximum jet density in the laboratory to scale to a maximum astrophysical jet density of ${\sim}$10${^{6}}$ cm${^{-3}}$. The constraints on the scaling discussed in sec. \ref{sec:3} give: 1 ns ${\rightarrow}$ 0.05 years, 50 eV ${\rightarrow}$ 3000 K, 50 T ${\rightarrow}$ 15 mG. The specific scaling applied in this case implies that the experimental flow, which lasts ${\sim}$200 ns, corresponds to an astrophysical outflow lasting ${\sim}$10 years. Such short times, when compared to the lifetimes of tens of thousands of years for protostellar outflows, may correspond to the ejection of a single ``clump'' as part of a more extended jet. The laboratory jet profiles to be scaled up are taken at a time approximately corresponding to the image in Fig. \ref{fig:7}b. The simulated astrophysical jet was evolved with the inclusion of cooling over ${\sim}$50 years on a Cartesian grid of $400\times10^{6}$ cells with a resolution of 2 AU. In these simulations the magnetic field is not included, and for the regime modelled here we would expect their inclusion to modify somewhat the dynamics. The flow dynamics is shown Fig. \ref{fig:9}. Initially the jet elongates because of the velocity variations imparted by the current-driven instability. The stretching of the jet is then followed by a rapid break up into smaller clumps which move at different velocities. The structure appearing in the large knot forming in the flow appears to be the result of Rayleigh-Taylor instabilities. In general, a single dense clump is produced by an ejection event like this, with the resulting outflow remaining well collimated over the propagation across ${\sim}$3000 AU. 

\section{Summary}
\label{sec:5}
Progress in high-energy density plasma experiments on lasers and z-pinch facilities has permitted in the last ten years to start investigating a range of ``large-scale'' astrophysical phenomena in the laboratory; extending the traditional domain of laboratory astrophysics beyond the work on micro-physics. Through careful design of the experiments, the plasma produced can be scaled to the astrophysical environment; allowing complex, intrinsically non-linear, three-dimensional phenomena to be accessed in a controlled manner. An important outcome being the validation of astrophysical codes on the laboratory data. Although work on astrophysical jets has been performed on both lasers and z-pinch facilities, we have focused here only on the studies of jets produced on the MAGPIE z-pinch facility. Two main ``types'' of jets were developed: hydrodynamic jets, to be used for propagation studies, and magnetohydrodynamic jets of interest to the launching phase. In general, there is some considerable control on the experiments: the initial condition can be partially modified, for example the density and magnetic field distributions; more complex physics, such as rotation, can be added, and different plasma condition can be produced, by modifying for example the cooling rates. Overall the combination of laboratory experiments and simulations can provide some very important insights on the physics of astrophysics, and as technology advances we can expect evermore exotic phenomena to be reproduced in the laboratory.

\subsubsection{Acknowledgements}
I would like to thank  C. Stehl\'{e} (Observatoire de Paris), S.V. Lebedev (Imperial College) and A. Frank (University of Rochester) for many useful discussions. This work was supported in part by the European Community's Marie Curie Actions-Human Resource and Mobility within the JETSET (Jet Simulations Experiments and Theory) network under contract RTN-CT-2004 005592. Access to the Marenostrum supercomputer, at the Barcelona Supercomputing Centre (Spain), was granted through the HPC-EUROPA project (RII3-CT-2003-506079), with the support of the European Community - Research Infrastructure Action under the FP6 "Structuring the European Research Area" Program. Finally, the author acknowledges the London e-Science Centre (LESC) for the provision of computational facilities and support.


%
 \bibliographystyle{plain}
 \bibliography{Ciardi_preprint}
%


%
\begin{table}
\centering
\caption{Characteristic conditions in laboratory (Z-pinch) and YSO jets}
\begin{tabular}{lll}
\hline\noalign{\smallskip}
 & Laboratory & YSO \\
\noalign{\smallskip}\hline\noalign{\smallskip}
Fluid velocity [km s$^{-1}$] & 100 - 400 & 100 - 500 \\
Density [g cm$^{-3}$] & $10^{-4}-10^{-6}$  & $10^{-18}-10^{-20}$ \\
Temperature [eV] & $5-200$ & $0.5-100$ \\
Magnetic field [G] & $10^{4}-10^{6}$ & $10^{-3}-10^{3}$ \\
Dynamical Age [ns] & $200-400$ & $10^{22}$ \\
Length [cm] & $2-4$ & $10^{17}$ \\
Radius [cm] & 0.5 & $10^{15}$ \\
Mass Flux [M$_\odot$ year$^{-1}$] & $10^{-33}$ & $10^{-7}-10^{-8}$ \\
Mean Ionisation & $5-10$ & $10^{-3}-1$ \\
Sound Speed [km s$^{-1}$] & $10^{3}-10^{4}$ & $10^{3}-10^{4}$ \\
Radiative Cooling Time [ns] & $4-40$ & $10^{18}$ \\
Mean Free Path [cm] & $10^{-5}$ & $10^{9}$ \\
Magnetic Diffusivity [cm$^2$ s$^{-1}$] & $10^{4}$ & $10^{8}$ \\
Kinematic Viscosity [cm$^2$ s$^{-1}$] & $10^{-3}-10$ & $10^{14}$ \\
Thermal Diffusivity [cm$^2$ s$^{-1}$] & $10^{3}-10^{6}$ & $10^{15}$ \\
\textbf{Mach number} & $5-40$ & $>> 5$ \\
\textbf{Re${_m}$} & $10-10^{3}$ & $> 10^{15}$ \\
\textbf{Re} & $> 10^{4}$ & $> 10^{8}$ \\
\textbf{Pe} & $50-10^{4}$ & $> 10^{7}$ \\
\textbf{Density Contrast} & $0.1-10$ & $> 1$ \\
\textbf{Cooling Parameter} & $0.01-10$ & $< 1$ \\
\textbf{Localization Parameter} & $< 10^{-4}$ & $< 10^{-6}$ \\
\textbf{Plasma-}${\beta}$ & $0.01-100$ & $0.01-100$ \\
\noalign{\smallskip}\hline
\end{tabular}
\label{tab:1}
\end{table}


\pagebreak 

\pagebreak 

\pagebreak 

\pagebreak 

\pagebreak 

\pagebreak 

\pagebreak 

\pagebreak 

\pagebreak 

\pagebreak 

\pagebreak 

\end{document}